# Repulsive Casimir forces between solid materials with high refractive index intervening liquids

P.J. van Zwol and G. Palasantzas

Department of Applied Physics, Materials innovation institute M2i and Zernike Institute for Advanced Materials, University of Groningen, 9747 AG Groningen, The Netherlands

**Abstract**

In order to explore repulsive Casimir/van der Waals forces between solid materials with liquid as the intervening medium, we analyze dielectric data for a wide range of materials as for example PTFE, polystyrene, silica and more than twenty liquids. Although significant variation in the dielectric data from different sources exist, we provide a scheme based on measured static dielectric constants, refractive indices, and applying Kramers Kronig (KK) consistency to dielectric data to create accurate dielectric functions at imaginary frequencies. The latter is necessary for more accurate force calculations via the Lifshitz theory allowing reliable predictions of repulsive Casimir forces.

Pacs mumbers: 78.68.+m, 03.70.+k, 85.85.+j, 12.20.Fv

# I. Introduction

Repulsive Casimir forces between two surfaces can be of high technological interest in low friction devices [1-4]. Measured repulsive Casimir forces have only been reported in a handful of papers [1-4]. As a matter of fact, repulsive Casimir forces between solids arise when the dielectric functions $\varepsilon$ of material surfaces 1 and 2 and an intervening liquid obey the relation $\varepsilon_1(i\zeta) > \varepsilon_{liquid}(i\zeta) > \varepsilon_2(i\zeta)$ over a wide frequency range $\zeta$. In most reports so far one of the surfaces is a metal, the other a low index material such as Teflon or silica, with an intervening non polar liquid having relatively high refractive index.

The major difficulty for predicting repulsive Casimir forces in liquids is that the calculated force itself is rather uncertain [5, 6]. This is because the dielectric function of most liquids in force calculations is not very different from the low index materials used in experimental systems [1-4]. Small variation in the dielectric data due to sample dependence or measurement uncertainty can easily lead to forces of opposite sign for one set of dielectric data compared to another for the same system. Contrary to solid materials, for very pure liquids we would not expect sample dependence of the dielectric function. This is because neither grains nor defects exist in a liquid nor the density varies significantly at atmospheric conditions. Therefore, it is relatively surprising that in the literature variations in dielectric data for liquids are often reported [6].

Furthermore, the more serious problem for all studies reported on repulsive Casimir forces is that the corresponding theory calculations were based on simple oscillator models, which were built from limited dielectric data (see [7] for the construction of oscillator models). Theoretical force predictions based on these models can easily be incorrect by an order of magnitude. Therefore, oscillator models are not suitable for accurate force calculations in solid-liquid systems [6]. On the other hand oscillator models were developed because precise knowledge of the VUV (vacuum ultraviolet) dielectric data for most substances was not available [7]. It is only in the last two decades that many dielectric data has become available, however, with large reported variation [6].

In order to overcome the problem of varying dielectric data in liquids and improve force calculations in terms of Lifshitz theory, we present a method to handle measured dielectric data that is based on the Kramers Kronig (KK) consistency of measured refractive index (n), extinction coefficient (k) or measured dielectric absorption, and the static dielectric constant $\varepsilon_0$. All dielectric data used here covers the major IR (infrared) and UV ranges providing therefore a reliable base for more accurate force calculations than in the past. Before proceeding we wish to point out that in essence, we combine the method behind the construction of oscillator models described in ref. [7] (for which the major concern is the determination of the fundamental frequencies of the oscillators), with frequency dependent dielectric data covering all the fundamental absorption frequencies.

## II. Dielectric data analysis: Application toTeflon

Polytetrafluoroethylene (PTFE or Teflon) is possibly the best well-known material for which Casimir repulsion is known [1-3]. However, the corresponding force measurements with this substance were compared to theory, which was based on simple 2-order oscillator models, since the complete dielectric function of PTFE was simply unknown. Here we were able to construct the dielectric function of PTFE from recent literature data (see appendix) [8-11] as Fig. 1 shows.

The VUV data gives the most important contribution to the dielectric function $\varepsilon(i\zeta)$ at imaginary frequencies as the inset in Fig. 1 inset shows. But, we have used only a similar molecule (poly(hexafluoro- 1.3-butadiene or PHFBD) to PTFE to obtain the dielectric function in this range because no other data was available. In general all fluoropolymers have very similar dielectric response with a first absorption peak at ~8 eV and a second one at ~20eV. In order to confirm the validity of using the PHFBD data further tests are necessary. For this purpose we use the KK transform,

$$n(i\zeta) = 1 + \frac{2}{\pi} \int_0^{+\infty} \frac{\omega k(\omega)}{\omega^2 + \zeta^2} d\omega \qquad (1)$$

with $\varepsilon(i\zeta) = n(i\zeta)^2$ the dielectric function at imaginary frequencies. The latter is important for force calculations in terms of the Lifshitz theory. If the KK transform is applied over all relevant frequency ranges then for $\zeta=0$ the function $\varepsilon(i\zeta)$ should reproduce the dielectric constant $\varepsilon_0$ [7]. In addition when a substance is transparent in the near IR and visible range (existence of well defined plateau in the $\varepsilon(i\zeta)$ vs. $\zeta$ plot; see Fig. 1), the KK transform over the UV and Xray frequency range yields for $\zeta=0$ the squared refractive index $n_0^2$ in the visible range (e.g., for example at the wavelength $\lambda=589.3$ nm). When IR absorption is negligible then $n_0^2 \approx \varepsilon_0$ holds [7]. The substances discussed in this work are all transparent in the near IR, visible, and near UV ranges. In addition, data for $\varepsilon_0$ and $n_0$ are known accurately (within a few percent accuracy) for almost any substance.

Therefore we are demanding that the less precise experimental data in the VUV regime is corrected in order to be KK consistent with the more accurate refractive index $n_o$ in the visible range or the dielectric constant $\varepsilon_0$ (if some IR absorption is present as in the case of PTFE; Fig. 1).

Although the choice of the wavelength $\lambda=589.3$ nm in the visible range where we obtain the value of $n_0$ might seems rather arbitrary, for most transparent substances it is located in the regime between the major IR and UV absorption peaks where also the value of $n(\omega)$ varies the least (well defined plateau in $\varepsilon(i\zeta)$ vs. $\zeta$; inset Fig. 1). Moreover, $n_0^2$ ($\lambda=589.3$ nm) is well measured and documented in the literature for almost any substance. For example for water $n_0$ varies from 1.32-1.34 for the wavelength range 1300 – 500 nm. The fact that $n_0^2(589.3nm) \approx \varepsilon_0$ within 1% for all non polar liquids with little IR absorption reflects this consideration (see also Table 1). Note at last that for the alkanes with negligible IR absorption (Table 1), the value of $n_0^2(589.3nm)$ agrees quite well (to within 2%) with the reported strength of $\varepsilon_0$ where Cauchy plots were used to determine $n_0$ [7]. In contrast to liquids with negligible IR absorption, for PTFE the value of $n_0^2(589.3nm)$ differs from $\varepsilon_0$ by ~15 % due to 2 major peaks in the IR range (Fig. 1).

Therefore, the dielectric function at imaginary frequencies should reproduce the $\varepsilon_0$ and $n_0^2$ values. If the measured data for the extinction coefficient k or the absorptive part of the dielectric function $\varepsilon(\omega)$ contain large uncertainties (~15% is not uncommon), then they can be multiplied by a factor within this uncertainty so that they reproduce the more precisely known value for $\varepsilon_0$ and $n_0^2$. The reason to use a renormalizing factor is that this is the simplest operation that can be applied on the data. Therefore we assume that the VUV data obtained from the literature does not contain large frequency dependent systematic errors, since in most studies these errors are not mentioned. On the other hand reported errors in the determination of the frequencies are generally very small, while reported errors in the absorption strength are larger. However by comparing data from different sources for the same substance (see below for example the case of Benzene) one can obtain an estimate for the accuracy of our assumption.

The procedure outlined above is particularly useful for liquids for which we would not expect atomic structure or density related variation in the dielectric data under constant atmospheric conditions. For solids a natural variation in the dielectric function due to varying structure (grains, defects, etc.) will also lead to a natural variation of $n_0^2$ and $\varepsilon_0$ [5, 6]. In all cases we only vary the VUV absorption since IR absorption is usually very small for liquids and known quite accurately. For PTFE a 15% correction was needed to reproduce the reported value of $\varepsilon_0=2.1$. By performing this correction the obtained value for $n_0$ was 1.77, which compares to reported values in literature of $n_0^2 \approx 1.74$-$1.85$ for PTFE surfaces (inset Fig. 1).

The method described above for creating dielectric functions at imaginary frequencies is far more accurate than creating simple oscillator models using only $\varepsilon_0$ and $n_0$ and some characteristic UV and IR frequencies of the substances under consideration. The problems with oscillator models become immediately clear when comparing them to real data (Fig 1). PTFE has at least two major absorption peaks in the UV range, while the simple models would only suggest one peak [7]. The model of Milling et al. [1] assumes the peak to be at 5 eV and that of Drummond et al. [12] at around 18 eV (compared to 8 and 20 eV as the measurement suggests). The model of [1] completely fails to describe the dielectric data. Although the model of ref [12] gives reasonable results, it also fails to describe the dielectric data in the far UV range.

In order to fit satisfactorily our derived data for $\varepsilon(i\zeta)$ at least five oscillators were necessary. The models of Drummond et al. [12] suggest a natural variation in dielectric data for different densities of PTFE. Unfortunately no other dielectric data are available from other sources for PTFE for comparison purposes. However, for polystyrene and silica multiple sets of data are available. Thus for the solids (except PTFE) we report two sets of data, while for liquids we will report only one. Note that for PTFE, a good approximation of sample dependence or density variation can be obtained by varying the dielectric strength at imaginary frequencies by ~5-10%.

**III. Cyclohexane and Benzene: Application of dielectric data analysis to liquids**

The obtained data for liquid cyclohexane (see also appendix) [11-15] are shown in Fig. 2. The measured data is taken from the solid state, where the density difference between solid and liquid cyclohexane is about 15% [15]. Since in the VUV domain atoms act as single absorbers, we can just apply a 15% correction to the data of solid cyclohexane in order to obtain the data for liquid, while any IR absorption is almost negligible. This is also reflected from the fact that $n_0^2$ and $\varepsilon_0$ are almost identical ($n_0^2 \approx \varepsilon_0 = 2.02$) [16]. Notably we had to apply an additional 8% correction to the VUV values of the reconstructed liquid data for $\varepsilon''$ in order to reproduce $n_0^2$ and $\varepsilon_0$ for liquid cyclohexane (which is within the reported experimental accuracy of ref [13]).

For Benzene we have two sets of data in the VUV range [13,17] (Fig. 3). Benzene has almost no IR absorption [18], which is reflected by the fact that $n_0^2 \approx \varepsilon_0$. The two different sets of UV data [13,17] lead to quite different dielectric strength at imaginary frequencies. When performing KK analysis on the data of ref [17] we obtained $n_0^2$=2.09, while the same procedure with the data of ref [13] gave $n_0^2$ =2.53. This is a difference of 20%, which can lead to very different forces [6]. In ref. [13] the estimated error is 10%, whereas the data was taken in the solid state suggesting that a correction is needed for the density as well. In ref [17] the reported accuracy in the absolute scale is 5%. However, in general, in the chemical literature (as shown in ref. [17]) a variation of ~10% for the obtained data is found [17]. Therefore, corrections to the reported values for $\varepsilon''$ and k in refs. [13] and [17] of 17% and 12% respectively were applied for KK consistency with the known values of $n_0^2=\varepsilon_0$=2.244 [16]. After correcting the data so that they reproduce the correct value of $n_0^2$ (and $\varepsilon_0$) the difference in dielectric functions at imaginary frequencies for refs [13] and [17] is less than 3% as the inset in Fig. 3 shows.

**IV. Halobenzenes: How to use similar molecules in dielectric data analysis**

For many high index liquids dielectric data over the required frequency range is not available. We will try here to show that when the dielectric function of a very similar molecule is known, then a good estimate of the dielectric function can still be made (similarly to the procedure applied for PTFE), and as an example we will apply this procedure to benzene derivatives for which force measurements exist. Note that the dielectric functions for the different benzene derivatives do not differ significantly from each other (Fig. 4a) [13]. All benzene derivatives are characterized by a thin absorption peak at ~7 eV and a broad one at ~20 eV, while they are transparent in the visible range.

We can construct the dielectric function of chlorobenzene at imaginary frequencies using the dielectric data for chlorobenzene [8, 13] and measured values for $n_0$. For comparison purposes, we will also create the dielectric function at imaginary frequencies of chlorobenzene using the dielectric data of pyridine and measured values of $n_0$ for chlorobenzene. The result of these calculations is shown in Fig. 4b. The difference is less than 2 % for the dielectric function at imaginary frequencies, suggesting clearly that this method works rather well.

Finally, we have used the VUV chlorobenzene dielectric data to construct the dielectric function of all the halobenzenes. Halobenzenes absorb strongly in the microwave regime for which we have no available dielectric data. Therefore we only used chlorobenzene data in combination with measured values for $n_0$ for the halobenzenes in applying the KK consistency (Table 1). IR absorption data was available for all halobenzenes [19-22], and toluene [23]. This information was also used for the construction of the dielectric function at imaginary frequencies in the range $10^{-3}$-$10^{2}$eV.

**V. Dielectric analysis for other liquids: Oscillator model representations**

Simple oscillator models (second or third order), which are built with dielectric data in a very limited frequency interval (or a few fixed frequencies), may not be very useful [6]. However in general the oscillator representation works well as a parameterization of any dielectric function at imaginary frequencies. We recall that the oscillator model is given by [7]

$$\varepsilon(i\zeta) = 1 + \sum_i \frac{C_i}{1+(\zeta/\omega_i)^2}. \qquad (1)$$

The coefficient $C_i$ is the oscillator strength at a given (resonance) frequency $\omega_i$. We have fitted these functions to the dielectric functions at imaginary frequencies as obtained with the method described previously. This is done because oscillator functions are convenient to use with Lifshitz theory to calculate the Casimir force. Since in our case they are obtained from measured dielectric data over a wide frequency interval, they should prove quite accurate. Oscillator model representations for many substances, which were constructed as proposed here, are shown in Table 1. Finally, the description of how we obtained the dielectric data for the remaining substances [24-40] shown in Table 1 is described in the appendix.

For most substances the valid frequency range of the presented oscillator models is $10^{-2}$-$10^{2}$ eV, and this is sufficient for finite temperature force calculations because the first IR Matsubara term is in this range. For some substances (such as silica and PTFE) the presented oscillator model is valid for any frequency. Recall that for solids we used two sets of dielectric data because of sample dependence [5, 6], and for liquids only one because we do not expect any sample dependence.

### VI. Theory calculations compared to force measurements

Measured repulsive Casimir forces have never been compared to Lifshitz theory calculations [41] based on measured dielectric data. Always for one or more substances simple (sometimes erroneous) 2 or 3 oscillator models were used [1-4, 6]. We will compare finite temperature force calculations using Lifshitz theory [41, 42], by implementation of the dielectric functions for liquids created with the method presented here, to force measurements [4, 42]. All force measurements were performed between gold and silica surfaces with different liquids [4,43].

For bromobenzene weak repulsion was measured, while for methanol and ethanol weak attraction (Fig. 5). The forces obtained with Lifshitz theory using the dielectric data from Table 1 (for the two sets of silica and the liquids) and the data for gold from ref. [5] are also shown. A variation of ~30% for the theory is observed due to sample dependence of silica surfaces. The calculations agree at the 30% level with the experiments for all cases. Considering the force variation of 30% due to variation in optical properties of solids, and the accuracy issues mentioned in ref. [6, 43], this is a good agreement. Note that the experimental force curves for ethanol and methanol do not appear to differ by ~30% as theory suggests, we attribute this to experimental uncertainty [43].

Note that in ref. [6] a variation of 100% was present for the silica-ethanol-gold system, since the reported variation in dielectric data for ethanol [6] was also taken into account. The whole purpose of the work presented here is to deal with such a variation (which we do not expect for liquids). As a result the improvement in theoretical accuracy is roughly a factor of three for these systems. Although for PTFE and gold surfaces almost any liquid leads to repulsion, this is not the case for silica and gold surfaces. Therefore silica provides an excellent test case, and a system for which, using the correct liquid, a sign switch of the force with distance can be expected.

**VII. Casimir force sign switch with distance: Is it observable?**

The smaller the difference between the dielectric function of the liquid and the low index solid, the larger the uncertainty in force becomes. However, the more likely a sign switch with distance [6]. This is shown in fig. 6 for silica and gold surfaces and a family of benzene liquids. The calculations were performed with two different sets of dielectric data for silica. For Fluorobenzene the force switches sign at 25 nm for one silica sample or it is attractive for the other silica sample. For Benzene it switches sign at 3 nm or it is attractive with unusual force scaling at 50 nm. For Chlorobenzene it switches sign at 3 nm or it is repulsive. Note that either the force switches sign at extremely small distances, or when it switches sign at larger distances it is very weak. Therefore a very sensitive force measurement device is needed for the detection of the sign switch.

We obtained similar results for PTFE and gold, in combination with low absorption liquids as water, methanol and pentane. In this case the force switches sign, being attractive below 10 nm. For ethanol we obtained repulsion (two times weaker than for cyclohexane). Therefore our results are also in agreement with those presented in ref [3]. For all other liquids the force was repulsive in case of PTFE and gold surfaces. For a slightly higher density PTFE sample, short range attraction may be observed for more liquids. For a slightly lower density PTFE sample even for water one may obtain divergent repulsion. However, we would like to note that water is notorious for capillary stiction, even in liquid environments [3, 44].

### VIII. The case of Casimir repulsion

There are not many combinations of materials for which Casimir repulsion is obtained. There are even less combinations for which this repulsion is strong. PTFE is the material of choice if we want to obtain strong Casimir repulsion because it has the smallest dielectric strength of any material. Stronger Casimir repulsion would be easier to measure, and the theoretical uncertainty would also be smaller since the dielectric contrast is larger [6]. Force calculations with some high index liquids between gold and PTFE are shown in Fig. 7. Compared to cyclohexane, using benzene and bromobenzene, already yields 3-4 times stronger repulsion, while Di-iodomethane gives even 10 times stronger repulsion. Since the latter liquid has one of the highest refractive indices of any pure liquid, much stronger repulsion will be very hard to obtain. For example liquids with refractive index $n_0 > 1.74$ have very undesirable other properties, and are often based on solutions with (highly toxic) solid components. In any case, Fig. 7 thus shows that using real liquids, Casimir repulsion can be varied within a factor roughly of 10 with a typical PTFE surface. Note that when using silica instead of PTFE, repulsion will be much weaker, or attractive. Thus it should be possible increase the repulsive force by a factor 5-10, compared to what has been measured to date [1-4]. Finally, note that van der Waals or Casimir repulsion in liquids is stable with respect to perturbations, it exists between macroscopic objects, and does not require dedicated geometry as it is required for obtaining repulsion in air/vacuum [45].

## IX. Conclusions

We have presented here measured dielectric data obtained from the literature for PTFE, polystyrene silica and more than twenty liquids. A scheme was provided to deal with the variation in dielectric data for liquids, which leads to high uncertainty in Casimir-Lifshitz force calculation. Using this scheme we have created dielectric data at imaginary frequencies for liquids, and we provided arguments why this data should be accurate to within a few percent error. Consequently we were able to provide more accurate force calculations for systems immersed in liquids. These force calculations with the new data are in good agreement with recent measurements in bromobenzene and alcohols. Besides cyclohexane we have suggested liquids for which much stronger Casimir repulsion exists. Dielectric data for simple oils was presented and this can be very useful for ultra low friction lubrication [3] using repulsive Casimir forces.

Force calculations with the new data suggest that a sign switch with distance will be very hard to measure due to the extremely small forces involved. Because of sample variation of solids it is suggested that a 'family' of similar liquids is required for the detection of this phenomenon, or the dielectric function of the solid should be accurately measured and an appropriate liquid is chosen. The dielectric data provided here should also be very useful for accurate prediction of contact angles [46], and more precise calculations for quantum torques [47].

## Acknowledgements

The research was carried out under Project No. MC3.05242 in the framework of the Strategic Research Program of the Materials Innovation Institute M2i (the former Netherlands Institute for Metals Research NIMR). The authors benefited from exchange of ideas by the ESF Research Network CASIMIR.

**Appendix**

*PTFE dielectric data:* The dielectric data in the Xray region >30eV can be modeled as in [8]. The vacuum UV data range 5-25eV is taken from a molecule very similar to PTFE (poly(hexafluoro- 1.3-butadiene)) [9]. In the visible and near IR Teflon is transparent and as a result absorption is negligible in this range. IR and THZ data were taken from refs [10] and [11].

*Polysterene and Silica:* Two sets of data for silica were obtained from ref [24]. For Polystyrene THZ and IR data was obtained from refs [25, 26], and the two different sets of UV data were obtained from refs [27, 28].

*(Cyclo)alkanes:* The procedure for liquid cyclohexane is as follows. Dielectric data for the XRay range >40eV is modeled [11]. Data in the range 10-40eV is taken from ref [13]. The IR range is taken from [14]. Particularly the errors in the VUV data of ref [13] are ±10%. This data is the most important for the construction of the dielectric function at imaginary frequencies.

The dielectric function of the other Cycloalkanes were all constructed from the data for cyclohexane and corrected to reproduce the corresponding $n_0^2$ and $\varepsilon_0$ values for a specific cycloalkane [16, 29]. For the simple Alkanes UV data in the range 7-200 eV was taken from ref [30]. The IR or THZ absorption was almost negligible for all (cyclo)alkanes, reflected by the fact that $n_0^2 \approx \varepsilon_0$ within the error bars [16]. The dielectric function for dodecane was estimated from $n_0^2$ and $\varepsilon_0$ and dielectric data of octane.

*Water and alcohols:* The dielectric data for water was obtained from Segelstein [31]. This data produces a too high value for $n_0^2$. Therefore, we applied a 15% correction to the absorption in the UV range (>5eV). While there are many other studies for water, we do not wish to provide a review for water here and we use only the data of Segelstein. Note that Segelsteins data [31] for k is also not perfectly KK consistent for n as it differs by 15%. For the Alcohols UV data was obtained from [6, 8, 32-35], and IR an THZ from [36, 37], again the UV data was made consistent with measured values for $n_0^2$ [16].

*Halo-alkanes:* For Methyl-Iodide dielectric data in the UV range 4-500 eV was found from [38], and IR data in the range 0.05-0.5eV [39]. A very similar liquid Methyl-Diiodide has one of the highest refractive indices ($n_0$=1.74) of any pure liquid while being not very toxic but slightly unstable under light. The dielectric function of Methyl-Diiodide was constructed using the IR data [39], and Methyl-Iodide UV data [38] were corrected with a factor to reproduce $n_0$=1.74. For tetrachloromethane data in the IR (0.02-0.5eV) and UV (5-200eV) ranges were obtained from ref. [22, 40]. The magnitude of the VUV absorption [40] was varied 15% to reproduce $\varepsilon_0$. This resulted in a value for $n_0$=1.449 as compared to the reported value of 1.461 [16].

*Viscous - oily liquids:* Finally we present data for simple highly viscous or oily liquids. These liquids are very interesting, since besides lubrication, repulsive Casimir forces will result in even lower or negligible friction with some substances [3]. One of these liquids is styrene. It is the building block of polystyrene and it has high refractive index. We have used the dielectric data of polystyrene and constructed the dielectric function of styrene to properly reproduce $n_0$ and $\varepsilon_0$. Repulsive forces even result for the silica-styrene-metal system. Thus styrene is extremely interesting for systems were glass and metal parts come into contact, which is not uncommon in micro/nano-mechanical devices. While it is not as strongly absorbing as styrene, glycerol has relatively high refractive index and it is often used in colloid physics. Compared to the simple alcohols glycerol ($C_3H_8O_3$) is structurally most similar to methanol ($CH_4O$). It also has a high value for $\varepsilon_0$ like methanol. Therefore, we will use the data of methanol and multiply with a factor to reproduce $n_0$=1.4697 for glycerol. For all simple alcohols the structure of the UV peaks does not change significantly apart from a factor, and IR absorption for liquids is weak. As a result we expect the force calculations, based on the constructed dielectric function at imaginary frequencies, to be quite precise like for all the other liquids.

**Figure Captions**

**Figure 1:** (Color online) (a) dielectric function at imaginary frequencies for PTFE. The thick black line is from data as described in text. Thin lines are oscillator models for low and high density teflon derivatives [12]. Red dashed lines are oscillator models from ref. [1]. The $n_0^2$ plateau is also shown (horizontal green) dashed line. (b) raw dielectric data for PTFE.

**Figure 2:** (Color online) Dielectric data for cyclohexane. The inset shows dielectric data at imaginary frequencies.

**Figure 3:** (Color online) Dielectric data for benzene from different sources. The dashed line was taken from ref [17] and solid line from ref. [13]. The data shown in the graph is already corrected so that it reproduces the correct value for $n_0^2$.

**Figure 4:** (Color online) (a) Dielectric functions from several benzene derivatives. (b) The dielectric function of chlorobenzene as constructed from dielectric data of Pyridine (dashed line), and Chlorobenzene (solid line) normalized using the indicated value of $n_0^2$.

**Figure 5:** (Color online) Force measurements between sphere and plate for silica and gold surfaces and the indicated liquids. Dots are experimental data from refs. [4, 40]. The lines represent theory using data from Table 1 and ref. [5]. We use two sets of dielectric data for silica. For bromobenzene the used sphere was 40 µm in diameter, while for ethanol and methanol a sphere with diameter 18 μm was used. Experimental force data for the bromobenzene system was digitized from a log-log plot [4], thus above 30nm, not all points are shown.

**Figure 6:** (Color online) Casimir forces between a silica sphere (diameter 18μm) and a gold surfaces immersed in various liquids as indicated. The force calculations were performed for two dielectric data sets for silica. Plots are shown for repulsive (upper part) and attractive (lower part) forces.

**Figure 7:** (Color online) Casimir forces between PTFE and gold surfaces between a sphere (diameter 18μm) and a plate immersed in various high index liquids. The inset shows dielectric data for PTFE and the liquids.

**Table 1:** Dielectric data for several substances. In case of solids the variation of $\varepsilon_0$ and $n_0$ is shown. For liquids the mean and standard deviation (as obtained from different sources) are shown at 298 K. Note that for liquids the variation in values is much lower than that of solids. The $\varepsilon_0$ and $n_0$ values are obtained from Landolt-Bornstein database averages [16]. The oscillator model parameters for the dielectric function at imaginary frequencies, valid in at least a range of $10^{-2}$-$10^2$ eV, are also shown. The first line shows the strength $C_i$ and the second line shows the frequency $\omega_i$ (eV). Note that for different substances a different number of oscillators is used. Furthermore the values of $C_i$ and $\omega_i$ are the result from a least square fit. They are as such a bit arbitrarily generated, and may not correspond to physical frequencies of the molecule. However, the fits agree always well with the fitted data (to within 1%)..

| Material | $\varepsilon_0$ | $n_0$ 589.3nm | Oscillator model values |
|---|---|---|---|
| **Low k Solids** | | | |
| PTFE | 2.1 | 1.32-1.36 | 9.30e-3 1.83e-2 1.39e-1 1.12e-1 1.95e-1 4.38e-1 1.06e-1 3.86e-2<br>3.00e-4 7.60e-3 5.57e-2 1.26e-1 6.71e+0 1.86e+1 4.21e+1 7.76e+1 |
| Silica (set 1) | 3.9-4.1 | 1.45-1.48 | 7.84e-1 2.03e-1 4.17e-1 3.93e-1 5.01e-2 8.20e-1 2.17e-1 5.50e-2<br>4.11e-2 1.12e-1 1.12e-1 1.11e-1 1.45e+1 1.70e+1 8.14e+0 9.16e+1 |
| Silica (set 2) | 3.9-4.1 | 1.45-1.48 | 1.19e+0 6.98e-2 1.35e-2 7.10e-1 1.80e-1 5.95e-1 2.27e-1 5.58e-2<br>5.47e-2 1.23e-2 5.74e-4 1.29e-1 9.10e+0 1.43e+1 2.31e+1 7.90e+1 |
| Polystyrene (set 1, 2008) | 2.4-2.5 | 1.55-1.59 | 1.21e-2 2.19e-2 1.79e-2 3.06e-2 3.03e-1 6.23e-1 3.25e-1 3.31e-2<br>1.00e-3 1.32e-2 3.88e+0 1.31e-1 5.99e+0 1.02e+1 1.88e+1 5.15e+1 |
| Polystyrene (set 2, 1977) | 2.4-2.5 | 1.55-1.59 | 3.12e-2 1.17e-2 2.17e-2 9.20e-3 2.93e-1 6.54e-1 4.17e-1 2.13e-2<br>1.18e-1 9.00e-4 1.19e-2 1.56e+0 6.12e+0 1.01e+1 2.02e+1 6.86e+1 |
| **Cycloalkanes** | | | |
| Cyclopentane | 1.96<br>() | 1.403<br>() | 1.51e-2 1.19e-1 4.79e-1 3.00e-1 5.07e-2<br>2.16e-1 8.09e+0 1.07e+1 1.79e+1 3.97e+1 |
| Cyclohexane | 2.01<br>0.02 | 1.424<br>0.002 | 1.51e-2 1.34e-1 5.17e-1 3.05e-1 4.65e-2<br>2.16e-1 8.03e+0 1.09e+1 1.84e+1 4.13e+1 |
| Cycloheptane | 2.08<br>() | 1.44<br>() | 1.52e-2 1.50e-1 5.80e-1 3.18e-1 1.81e-2<br>2.18e-1 8.26e+0 1.09e+1 2.04e+1 6.02e+1 |
| Cyclooctane | 2.12<br>() | 1.454<br>() | 1.47e-2 3.75e-1 5.45e-1 1.71e-1 1.57e-2<br>2.08e-1 8.67e+0 1.36e+1 2.36e+1 6.61e+1 |
| **Benzene derivatives** | | | |
| Benzene | 2.273<br>0.009 | 1.498<br>4e-4 | 2.32e-2 6.99e-3 8.51e-2 2.57e-1 6.59e-1 2.26e-1 1.43e-2<br>8.76e-1 6.71e+0 4.48e+0 1.70e+1 8.48e+0 2.33e+1 7.01e+1 |
| FluoroBenzene | 5.3<br>0.1 | 1.4629<br>(2e-6) | 6.14e-2 2.38e-2 4.94e-1 5.29e-1 9.63e-2 1.24e-2<br>1.29e-1 5.52e-2 7.04e+0 1.53e+1 2.68e+1 8.91e+1 |
| ChloroBenzene | 5.75<br>0.23 | 1.522<br>0.001 | 4.92e-2 2.64e-2 3.77e-1 5.89e-1 3.19e-1 3.18e-2<br>7.89e-2 1.46e-1 6.27e+0 1.16e+1 2.03e+1 5.27e+1 |
| BromoBenzene | 5.37<br>0.04 | 1.558<br>0.001 | 5.44e-2 1.84e-2 4.75e-2 5.32e-1 6.45e-1 2.40e-1 9.27e-3<br>5.02e-3 3.09e-2 1.11e-1 6.75e+0 1.33e+1 2.40e+1 9.99e+1 |
| IodoBenzene | 4.6<br>() | 1.62<br>() | 4.08e-3 7.98e-2 7.98e-3 4.20e-1 7.40e-1 4.46e-1 1.26e-2<br>2.60e-2 9.40e-2 1.88e+0 6.37e+0 1.09e+1 2.14e+1 8.86e+1 |
| Toluene | 2.395<br>0.02 | 1.494<br>5e-4 | 5.61e-3 6.97e-2 8.07e-3 5.15e-1 5.74e-1 9.91e-3 1.18e-1 1.01e-2<br>3.40e-2 9.97e-2 1.15e+0 7.23e+0 1.50e+1 2.08e+1 2.59e+1 7.65e+1 |
| **Alkanes** | | | |
| Pentane | 1.831<br>0.02 | 1.355<br>() | 1.91e-2 1.03e-1 4.16e-1 2.39e-1 5.85e-2<br>2.38e-1 7.98e+0 1.32e+1 1.79e+1 3.86e+1 |
| Hexane | 1.887<br>0.004 | 1.3727<br>6e-4 | 1.86e-2 1.09e-1 4.43e-1 2.82e-1 3.25e-2<br>2.31e-1 7.90e+0 1.29e+1 1.93e+1 4.81e+1 |
| Heptane | 1.914<br>0.01 | 1.3853<br>7e-4 | 1.89e-2 1.12e-1 4.59e-1 2.89e-1 3.43e-2<br>2.31e-1 7.94e+0 1.28e+1 1.92e+1 4.79e+1 |
| Octane | 1.934<br>0.02 | 1.3951<br>4e-4 | 1.94e-2 1.14e-1 4.72e-1 2.92e-1 3.57e-2<br>2.34e-1 7.52e+0 1.30e+1 1.87e+1 4.92e+1 |
| Dodecane | 2.014<br>0.013 | 1.420<br>0.003 | 1.71e-2 1.15e-1 5.66e-1 2.68e-1 4.65e-2<br>2.17e-1 7.55e+0 1.30e+1 1.90e+1 4.45e+1 |
| **Alcohols** | | | |
| Water | 78.7<br>0.4 | 1.3325<br>4e-4 | 1.43e+0 9.74e+0 2.16e+0 5.32e-1 3.89e-1 2.65e-1 1.36e-1<br>2.29e-2 8.77e-4 4.93e-3 1.03e-1 9.50e+0 2.09e+1 2.64e+1 |
| Methanol | 32.9<br>0.4 | 1.3266<br>4e-4 | 4.38e-1 3.24e-1 1.38e-1 1.20e-1 4.45e-1 2.52e-1 6.49e-2<br>6.10e-3 1.93e-2 1.02e-1 4.35e-1 1.08e+1 2.01e+1 3.83e+1 |
| Ethanol | 24.8<br>0.4 | 1.3595<br>5e-4 | 9.57e-1 1.62e+0 1.40e-1 1.26e-1 4.16e-1 2.44e-1 7.10e-2<br>1.62e-3 4.32e-3 1.12e-1 6.87e+0 1.52e+1 1.56e+1 4.38e+1 |
| Propanol | 20.2<br>0.5 | 1.3834<br>6e-4 | 4.17e-1 3.74e-1 1.87e-1 1.47e-1 4.89e-1 3.06e-1 1.22e-1<br>1.97e-4 7.77e-3 3.75e-2 3.41e-1 1.01e+1 1.75e+1 3.17e+1 |
| Butanol | 17.4<br>0.2 | 1.3974<br>6e-4 | 5.49e-1 3.14e-1 1.31e-1 1.10e-1 5.44e-1 2.79e-1 2.72e-2<br>2.15e-4 9.58e-3 8.26e-2 5.83e+0 1.25e+1 2.04e+1 6.16e+1 |
| **Halo-alkanes** | | | |
| Iodomethane | 6.89<br>() | 1.532<br>() | 1.70e-2 7.29e-1 2.63e-1 2.98e-1 4.79e-2<br>9.61e-2 8.01e+0 1.41e+1 1.63e+1 6.76e+1 |
| Diiodomethane | 5.318<br>0.002 | 1.74<br>() | 1.94e-2 9.25e-1 9.22e-1 8.59e-2 9.61e-2<br>9.38e-2 7.46e+0 1.33e+1 5.85e+1 1.37e+1 |
| Tetrachloro-<br>methane | 2.216<br>() | 1.461<br>() | 1.22e-1 2.85e-1 5.63e-1 2.36e-1 6.79e-3<br>9.53e-2 8.42e+0 1.34e+1 2.21e+1 8.74e+1 |
| **Oily liquids** | | | |
| Glycerol | 42.4<br>1.4 | 1.4697<br>() | 5.80e-1 3.23e-1 2.03e-1 7.68e-1 3.70e-1 2.20e-2<br>6.94e-3 3.04e-2 2.91e-1 1.07e+1 2.39e+1 5.04e+1 |
| Styrene | 2.47<br>() | 1.544<br>() | 1.31e-2 2.30e-2 3.08e-2 4.07e-1 6.04e-1 3.34e-1 3.89e-2<br>1.08e-3 1.50e-2 1.45e-1 6.24e+0 1.15e+1 1.98e+1 5.15e+1 |
| | | | |

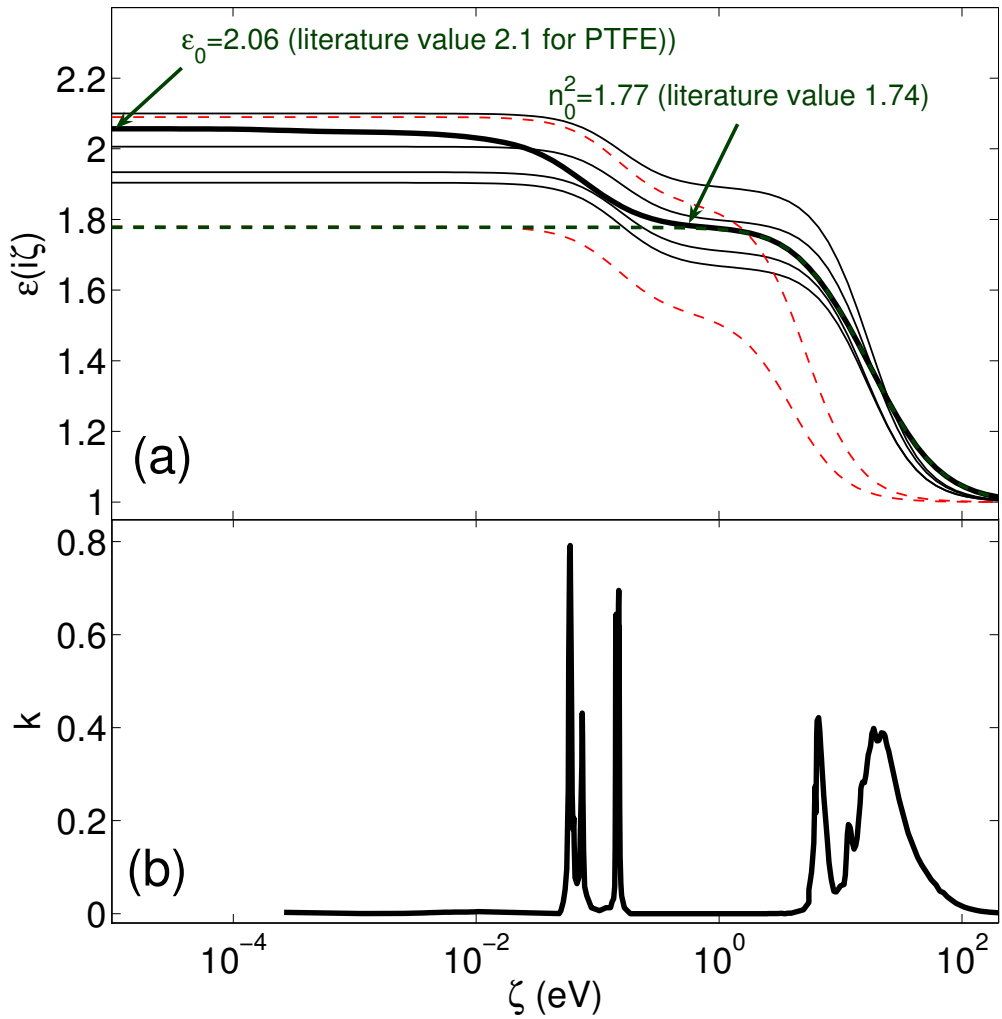

Figure 1

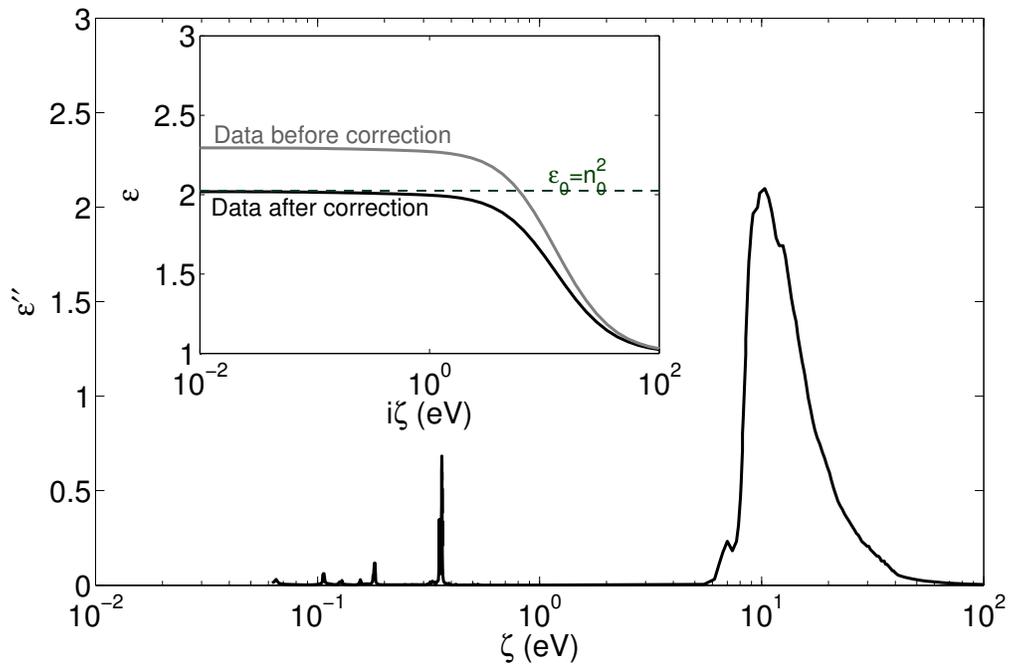

Figure 2

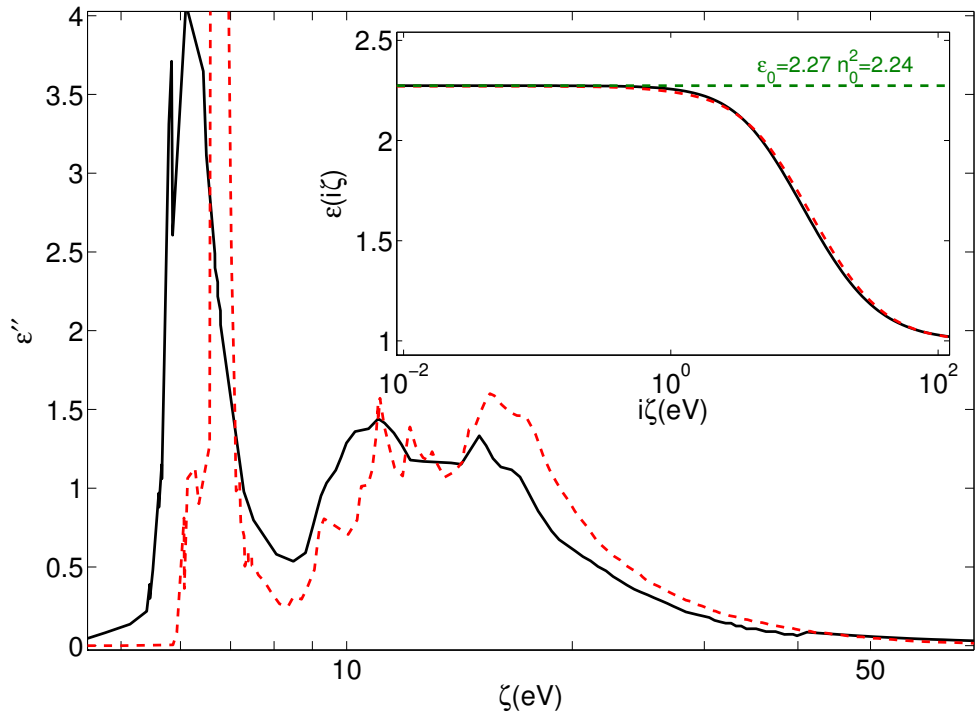

Figure 3

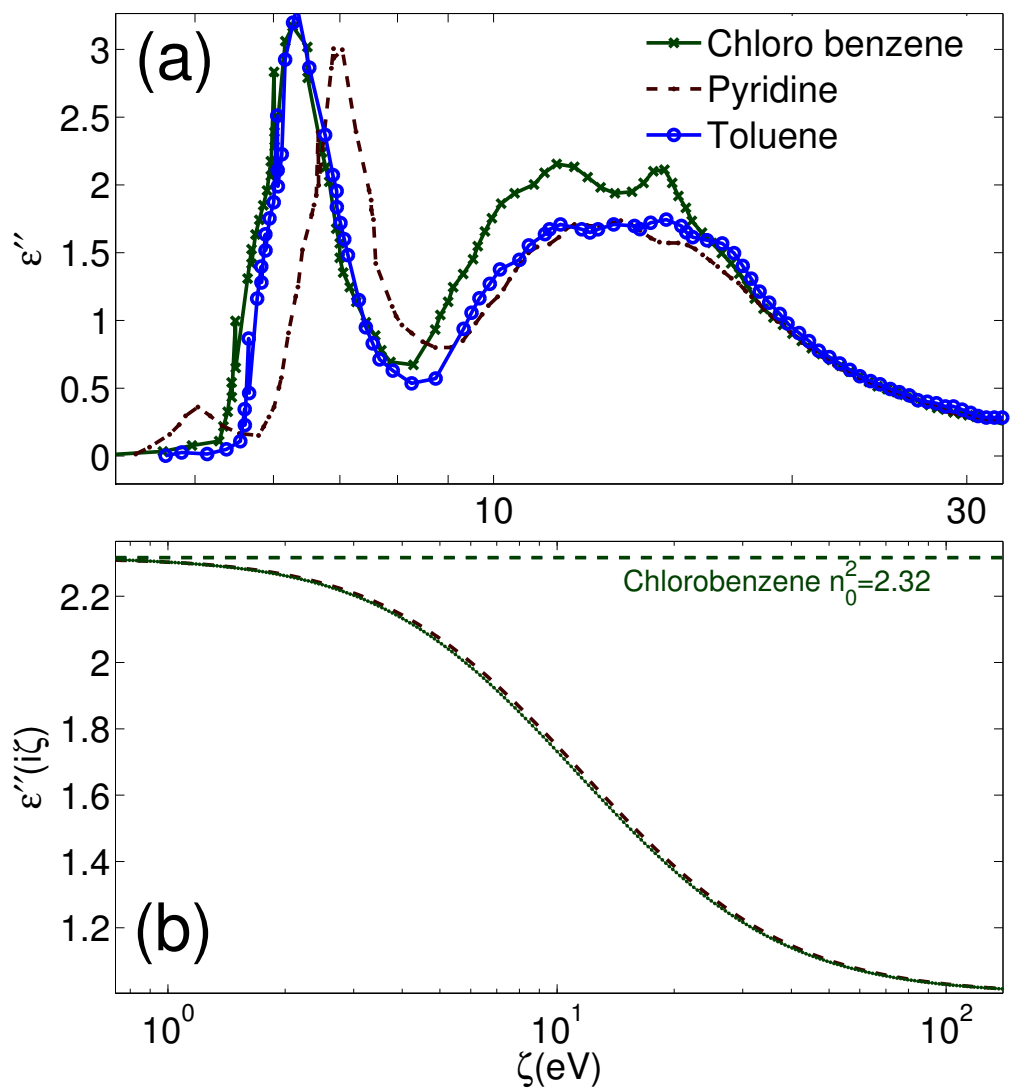

**Figure 4**

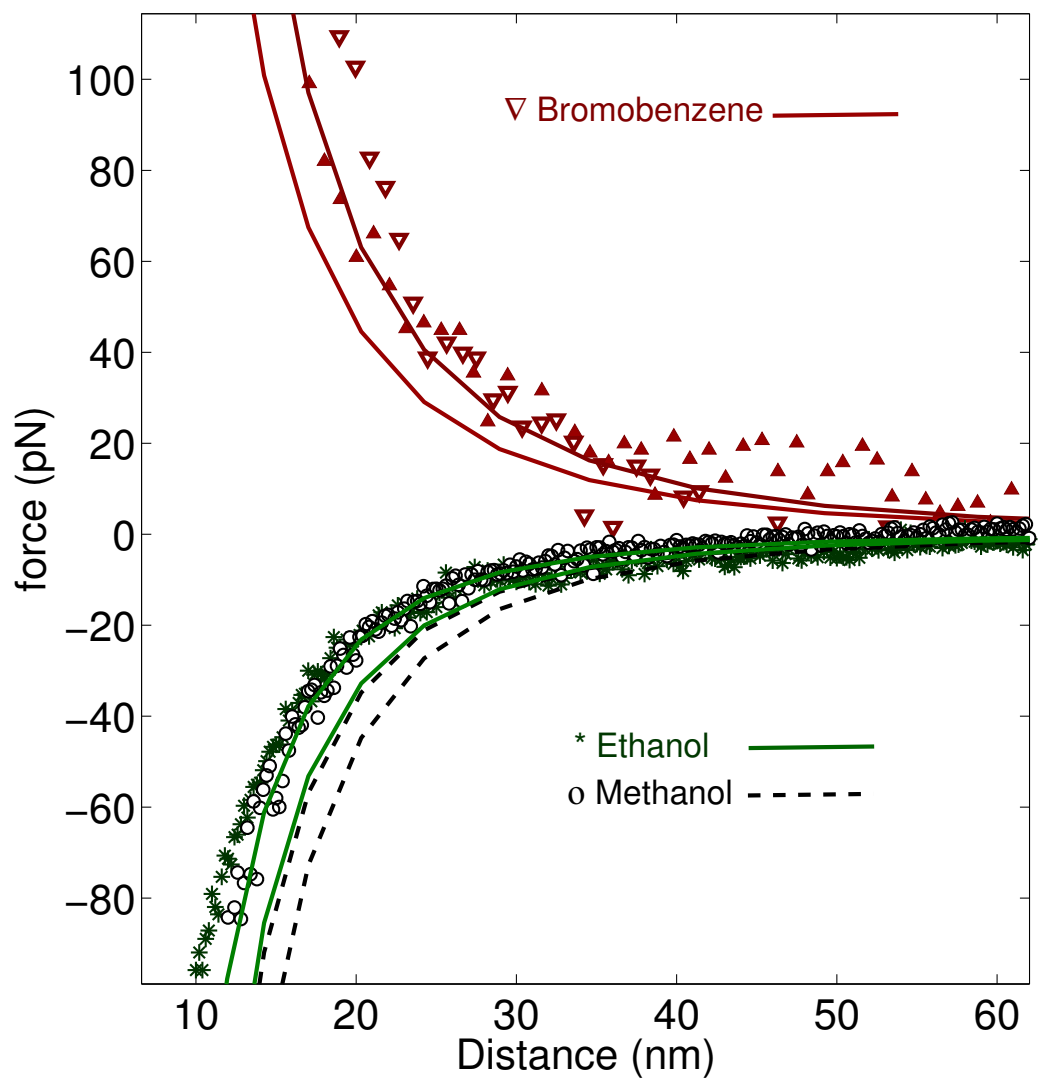

**Figure 5**

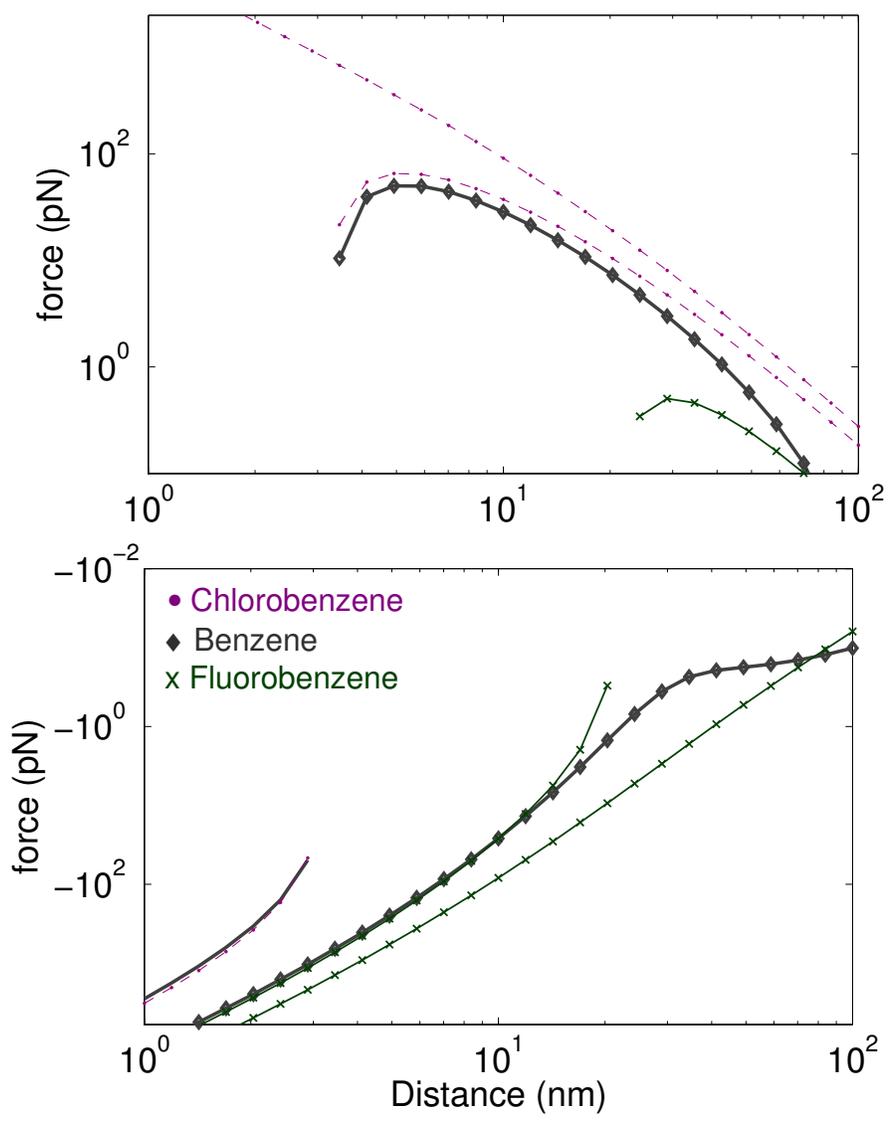

**Figure 6**

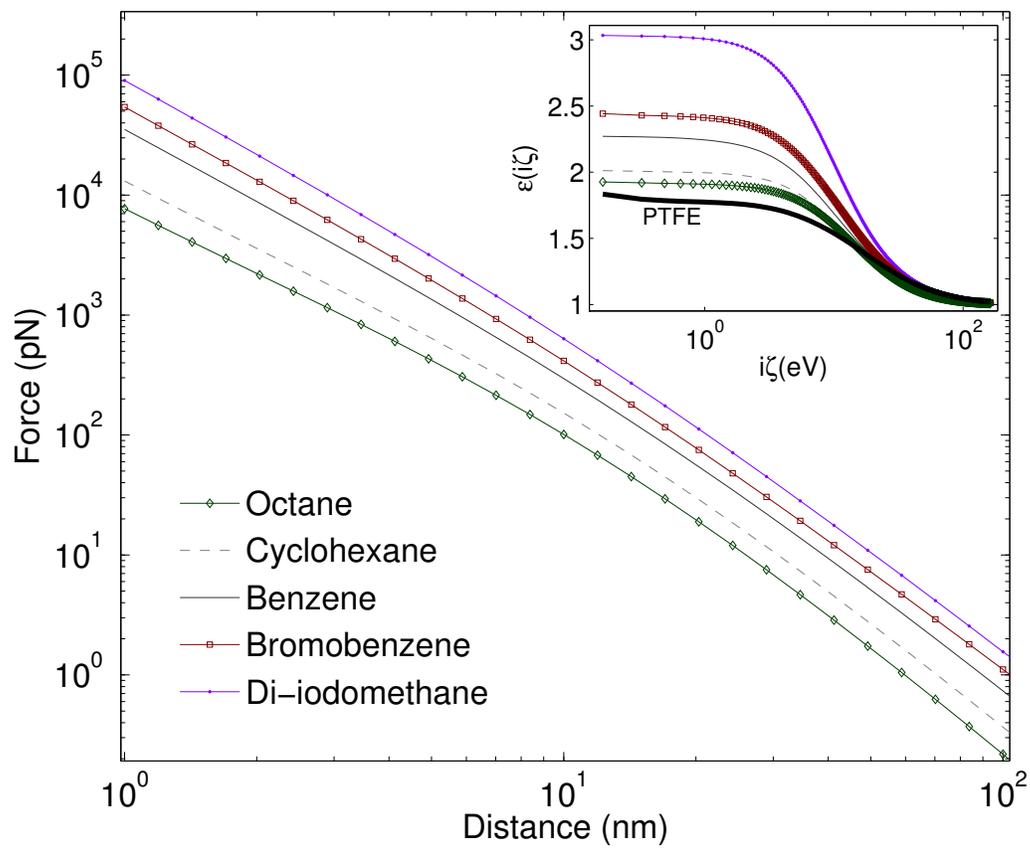

**Figure 7**